\documentclass[aps,prd,10pt,twocolumn,preprintnumbers,superscriptaddress,nofootinbib,letterpaper,longbibliography]{revtex4-2}

\usepackage{hyperref}
\usepackage{graphicx}
\usepackage{amsfonts,amsmath,amssymb,bm}
\usepackage{array}
\usepackage[utf8]{inputenc}
\usepackage{hyperref}
\usepackage{slashed}
\usepackage{xcolor}
\hypersetup{
  colorlinks  = true,
  urlcolor    = blue,
  linkcolor   = blue,
  citecolor   = red
}
\newcommand{\ehadvis}{$E_{\mathrm{had}}^{\mathrm{vis}}$\;}
\newcommand{\enureco}{$E_{\nu}^{\mathrm{reco}}$\;}
\newcommand{\enutrue}{$E_{\nu}^{\mathrm{true}}$\;}
\newcommand{\qreco}{$|\vec{q}\,|^{\mathrm{reco}}$\;}

\newcommand{\orcid}[1]
{\begingroup
  \hypersetup{hidelinks}\href{https://orcid.org/#1}{\includegraphics[width=9pt]{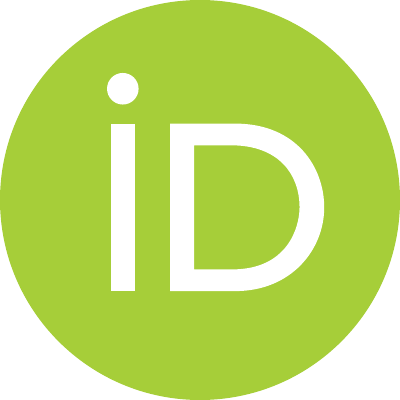}
} \endgroup}

\newcolumntype{P}[1]{>{\centering\arraybackslash}p{#1}}

\def\Fermilab{Theoretical Physics Department, Fermilab, P.O. Box 500, Batavia, IL 60510, USA}
\def\Irvine{Department of Physics and Astronomy, University of California, Irvine, CA 92697}

\begin{document}


\title{Neutrino-Nucleus Cross Section Impacts on Neutrino Oscillation Measurements}

\author{Nina M. Coyle \orcid{0000-0002-7440-496X}}
\email{ncoyle@uci.edu}
\affiliation{\Irvine}

\author{Shirley Weishi Li \orcid{0000-0002-2157-8982}}
\email{shirley.li@uci.edu}
\affiliation{\Irvine}

\author{Pedro A. N. Machado \orcid{0000-0002-9118-7354}}
\email{pmachado@fnal.gov}
\affiliation{\Fermilab}

\date{February 26, 2025}

\begin{abstract}
The challenges in neutrino-nucleus cross section modeling and its impact on neutrino oscillation experiments are widely recognized. However, a comprehensive and theoretically robust estimation of cross section uncertainties has been lacking, and few studies have quantitatively examined their impact on oscillation measurements. In this work, we evaluate the effect of cross section uncertainties on oscillation parameters using setups inspired by NOvA and DUNE. To characterize these uncertainties, we adopt multiple neutrino-nucleus event generators and simulate a realistic experimental procedure that incorporates near-detector data and near-to-far-detector extrapolation.
Our results confirm that cross section uncertainties cannot significantly bias oscillation results in current statistics-dominated experiments like NOvA. 
However, they could lead to substantial bias for future systematics-dominated experiments like DUNE, even when near-detector data are employed to mitigate uncertainties.
These findings underscore the need for further studies on the quantitative impacts of cross section modeling, improved strategies to utilize near-detector data and the PRISM concept, and more robust cross section models to optimize the success of future experiments.
\end{abstract}

\preprint{FERMILAB-PUB-25-0001-T, UCI-HEP-TR-2025-01}

\maketitle


\section{Introduction}

As neutrino physics enters a precision era with future experiments like DUNE, Hyper-Kamiokande, and JUNO~\cite{JUNO:2015zny,Hyper-Kamiokande:2018ofw,DUNE:2020ypp}, a rich physics program emerges for exploration, alongside an increasing need to understand neutrino-nucleus interaction physics~\cite{Mosel:2016cwa, NuSTEC:2017hzk, Ruso:2022qes,Coyle:2022bwa}. Long-baseline experiments such as DUNE and Hyper-Kamiokande, which utilize broadband neutrino beams, require inferring the neutrino energy of individual events from observed final states. This process relies on cross section models that predict the amount of missing energy due to factors like escaping neutrons or charged particles below detection thresholds. However, in the GeV energy range of these neutrino beams, the nuclear degree of freedom involves partons, nucleons, and baryonic resonances, making first-principle calculations of cross sections currently unfeasible~\cite{Ruso:2022qes}.

Given this complexity, experiments must adopt pragmatic approaches to model neutrino-nucleus interactions. To this end, they employ neutrino event generators~\cite{Casper:2002sd, Andreopoulos:2009rq, GENIE:2021npt, Buss:2011mx, Golan:2012wx, Bohlen:2014buj, Stowell:2016jfr, Gardiner:2021qfr, Hayato:2021heg, CLAS:2021neh, Isaacson:2022cwh}. These generators compute cross sections across different channels, each corresponding to a specific type of hadronic degree of freedom: nucleons for quasi-elastic scattering, nucleon pairs for meson exchange current, baryonic resonances for resonance production, and partons for deep-inelastic scattering (which may also include shallow inelastic scattering).  
All these channels are subject to nuclear effects, such as final-state interactions~\cite{Bertini:1963zzc, Buss:2011mx, Cugnon:1980zz, Bertsch:1984gb, Bertsch:1988ik, Cugnon:1996xf, Boudard:2002yn, Sawada:2012hk, Uozumi:2012fm, Golan:2012wx, Rocco:2020jlx, Isaacson:2020wlx, Dytman:2021ohr, Gonzalez-Rosa:2023aim}. While the models implemented in event generators provide reasonable estimates of cross sections, quantitative studies have revealed their limitations in accurately reproducing neutrino-nucleus and electron-nucleus scattering data~\cite{Meyer:2016oeg, MINERvA:2019rhx, Ankowski:2019mfd, Ankowski:2020qbe, MINERvA:2020anu, MINERvA:2020zzv, Borah:2020gte, T2K:2020jav, T2K:2021naz, Tomalak:2021qrg, MINERvA:2021owq, NOvA:2021eqi, Mosel:2023zek, MicroBooNE:2024zwf}.

To address cross section mismodeling and uncertainties, long-baseline experiments employ an additional tool by leveraging near detectors (ND). These detectors are placed sufficiently close to the neutrino beam source, where no measurable oscillation effects are expected. This proximity allows ND data to serve as a validation sample for refining both beam simulations and cross section models. When predictions from the default generator models deviate from ND data, experiments frequently perform ND tuning, adjusting the generator models to match the observed data. 
The resulting tuned generator is then used to predict oscillations in the far detector (FD).

Despite recent intensive efforts in neutrino cross section studies~\cite{Mosel:2016cwa, NuSTEC:2017hzk,Kronfeld:2019nfb,Meyer:2022mix,Ruso:2022qes}, including theoretical calculations and experimental measurements, few studies have investigated how cross section mis-modeling and uncertainties impact oscillation measurements. While existing oscillation measurements published by experimental collaborations account for cross section uncertainties and assign associated systematic errors, the full analysis details are not publicly available. Moreover, these analyses often include experimental effects that ideally should be disentangled from cross section physics. 
Although future experiments like DUNE and Hyper-Kamiokande have produced sensitivity studies using early detector simulation and cross-section models, neither has published an analysis with an uncertainty model sufficient for analyzing real data. 
While Refs.~\cite{Ankowski:2015kya, Ankowski:2016jdd, Ankowski:2017uvv, Nagu:2019uco, Devi:2022zfh} were the first to explore cross section uncertainties and their propagation to oscillation parameters in neutrino experiments, they only consider the FD.

There is an urgent need to answer the following question: given current cross section uncertainties, can we achieve the designed {\it accuracy} on oscillation parameters in next-generation experiments? Our paper aims to address this question. We characterize current cross section uncertainties by using multiple event generators: GENIE~\cite{Andreopoulos:2009rq,GENIE:2021npt}, GiBUU~\cite{Buss:2011mx}, and NuWro~\cite{Golan:2012wx}. For the utilization of ND data, we closely follow NOvA's analysis strategy as an instructive example, including ND tuning and near-to-far-detector extrapolation.

The paper is organized as follows. In Sec.~\ref{sec:methods}, we describe our analysis procedure and methodology, as well as the results of performing this analysis for the NOvA oscillation setup. In Sec.~\ref{sec:DUNE}, we present results for the DUNE oscillation analysis. Finally, we discuss our conclusions in Sec.~\ref{sec:conclusions}.

\section{Methods}
\label{sec:methods}

Our aim is to select an analysis strategy that closely approximates DUNE's future approach. 
DUNE is currently in the construction phase, and experiments like SBND~\cite{MicroBooNE:2015bmn} are actively collecting $\nu$-Ar cross section data that may inform DUNE's cross section model. Consequently, the full DUNE analysis procedure and cross section mitigation strategy have not yet been published. Given these circumstances, our analysis adopts the existing NOvA procedure~\cite{NOvA:2020rbg, NOvA:2021nfi, osti_1827395} as a proxy, leveraging the similarities between the energy ranges of the two experiments. 
It is important to note that this approach is not tailored to DUNE's specific needs and design, nor does it utilize DUNE-PRISM~\cite{DUNE:2021tad}. 
As such, we expect our findings to represent a conservative estimate of DUNE's potential accuracy. 
The actual DUNE oscillation analysis, when finalized, may achieve significant improvements beyond our current projections. 
In what follows, we describe the details of our analysis for both the NOvA and DUNE experiments.

First, to characterize the current theoretical uncertainties in cross section modeling, we face the challenge of lacking robust uncertainty estimates.
To address this, we employ three different event generators: GENIE (version 3.04.00 with G18\_02a\_00\_000 tune), NuWro (version 21.09.2), and GiBUU (release 2023). We treat the differences between these generators as a proxy for cross section uncertainties. While this approach is not entirely accurate---it may underestimate or overestimate true uncertainties---it provides a workable framework for our analysis. Specifically, we simulate mock data using GENIE, assuming it represents the true cross section model. We then conduct the analysis using either NuWro or GiBUU, simulating a scenario where these are the available generators. This method allows us to assess the impact of potential discrepancies between the assumed cross section model and the one used in the analysis.

Second, because the impact of cross section calculations arises primarily in the predicted relation between true and reconstructed neutrino energies, as will be discussed in Sec.~\ref{sec:FD prediction}, we must consider detector effects and reconstruction capabilities. We employ calorimetric energy reconstruction, in which the incoming neutrino energy is reconstructed as the sum of the total energy of outgoing lepton and mesons, plus the kinetic energy of visible hadrons.
For NOvA, we include a 30\% smearing on hadronic final-state particles and 3\% on leptons~\cite{NOvA:2020rbg}. 
We include proton quenching effects with Birks' constant and $dE/dx$ values taken from Ref.~\cite{Awe:2020xau} and NIST Standard Reference Database 124, respectively~\cite{NISTdatabase}; a proton kinetic energy threshold of 50 MeV; and a pion kinetic energy threshold of 20 MeV. 
Neutrons are considered invisible for NOvA. 
These thresholds are choices that enable us to approximately reproduce the NOvA visible hadronic energy spectrum.
For DUNE, we employ the resolutions and kinematic thresholds listed in Table 3.3 of Ref.~\cite{DUNE:2015lol} along with proton quenching effects. 
As described in the caption of that table, we consider 60\% of post-smearing neutron energy to be visible, and include a 10\% probability that a neutron with momentum less than 1 GeV will be missed by reconstruction techniques (see also Ref.~\cite{Friedland:2018vry}). 

In the following subsections, we provide a detailed description of NOvA's primary mitigation strategies---ND tuning and near-to-far-detector extrapolation---which we adopt in our analysis.

\subsection{Near-detector tuning procedure}

\begin{figure*}
    \centering    
    \includegraphics[width=0.95\textwidth]{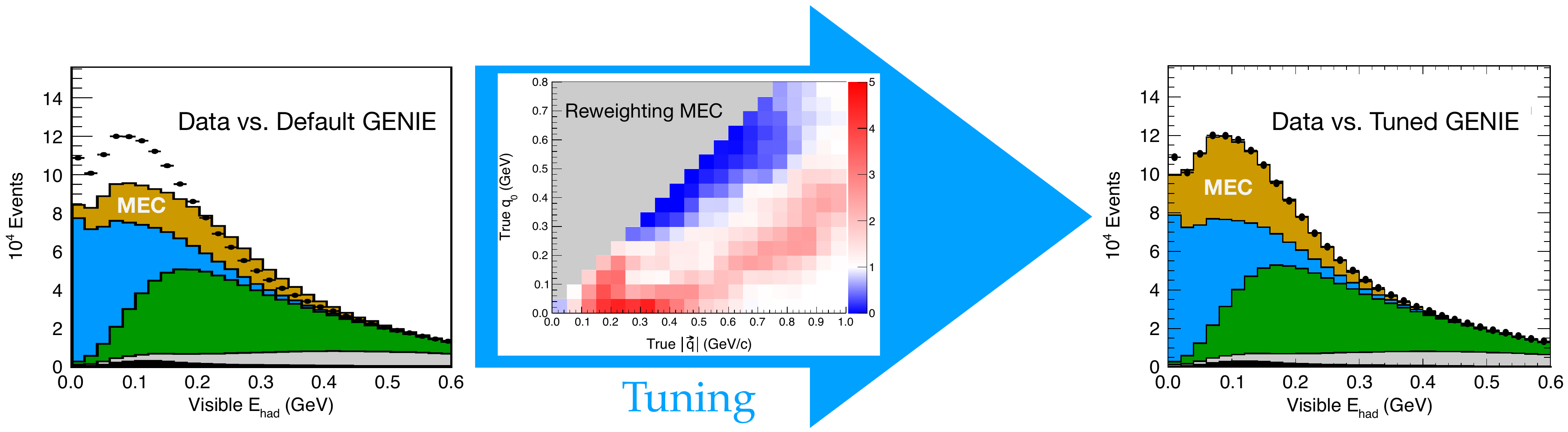}
    \caption{Illustration of NOvA ND tuning procedure; figure taken from Ref.~\cite{Coyle:2022bwa}, modified from Ref.~\cite{NOvA:2020rbg}.}
    \label{fig:NDtune}
\end{figure*}

Experiments frequently observe significant discrepancies between measured neutrino spectra and baseline generator predictions in the ND~\cite{MINERvA:2019kfr,NOvA:2020rbg,T2K:2023smv}. To address this, experiments utilize ND data to adjust the models implemented in event generators. The resulting modified generator is then employed in the unfolding and smearing procedures. It is important to note that the precise tuning method varies widely between experiments, and the resulting cross section model may not be universally applicable across different experiments or analyses.

NOvA employs the GENIE  event generator for their analysis. There are two primary pieces of the NOvA tune: adjustments to the base generator based on data from previous experiments, and a subsequent tune of the updated generator to NOvA ND data~\cite{NOvA:2020rbg}. 
Examples of the first step include adjusting the axial mass from GENIE default value 0.99 GeV to 1.04 GeV, based on a re-analysis of the deuterium bubble chamber data~\cite{Meyer:2016oeg}. We do not implement these initial modifications because we are using multiple different generators as a proxy for the possible impacts of cross section modeling. 
Regardless, significant discrepancies remain between the simulated and measured ND spectra even after these changes.

We implement only the tuning where the generator models are adjusted based on NOvA ND data, illustrated in Fig.~\ref{fig:NDtune}. While this precise approach is specific to the NOvA experiment, it provides an example of tuning from which we can take away some lessons and may influence the approach for DUNE. The tuning procedure involves the following:
\begin{itemize}
    \item The simulated events are binned in a two-dimensional plane of true energy transfer $q^0$ and true three-momentum transfer $|\vec{q}\,|$ to the hadronic system.
    \item Weights are applied independently to the meson exchange current (MEC) cross section in each bin of the two-dimensional true plane.
    \item The reweighted distribution in the true plane is migrated into the reconstructed \ehadvis and \qreco plane.
    \item The simulation is compared to data in the reconstructed plane, with the tune weights in the true plane iteratively updated to fit the data in the reconstructed plane.
\end{itemize}
The fit in this procedure employs a $\chi^2$ fit with statistical uncertainties for each bin in the (\qreco,\ehadvis) plane; the tune weights themselves are included as Gaussian penalty terms with an uncertainty of 100\%. In the experimental oscillation analyses, systematic uncertainties arising from cross section modeling are identified by adjusting the relevant GENIE inputs within their uncertainties. The uncertainties arising specifically from the ND MEC tuning procedure are determined by performing two different versions of the tune: one in which the QE contribution is enhanced by 1$\sigma$ before the ND tune, and another in which the RES contribution is instead enhanced. The spread of the resulting tuned spectra provides a systematic uncertainty for the MEC part of the ND tune.

Regarding the reweighting, it is important to note that the cross section adjustment is applied only to the MEC channel. 
This approach is based on the assumption that other channels---quasi-elastic, resonance production, and deep-inelastic scattering---have significantly less cross section mismodeling given the initial adjustments made before fitting to the NOvA ND data.
While the validity of this assumption in current cross section models is uncertain, our analysis suggests that this particular choice does not significantly impact the tuned results. We will discuss this further in our mock DUNE analysis.

\begin{figure*}[t]
    \centering    
    \includegraphics[width=0.8\textwidth]{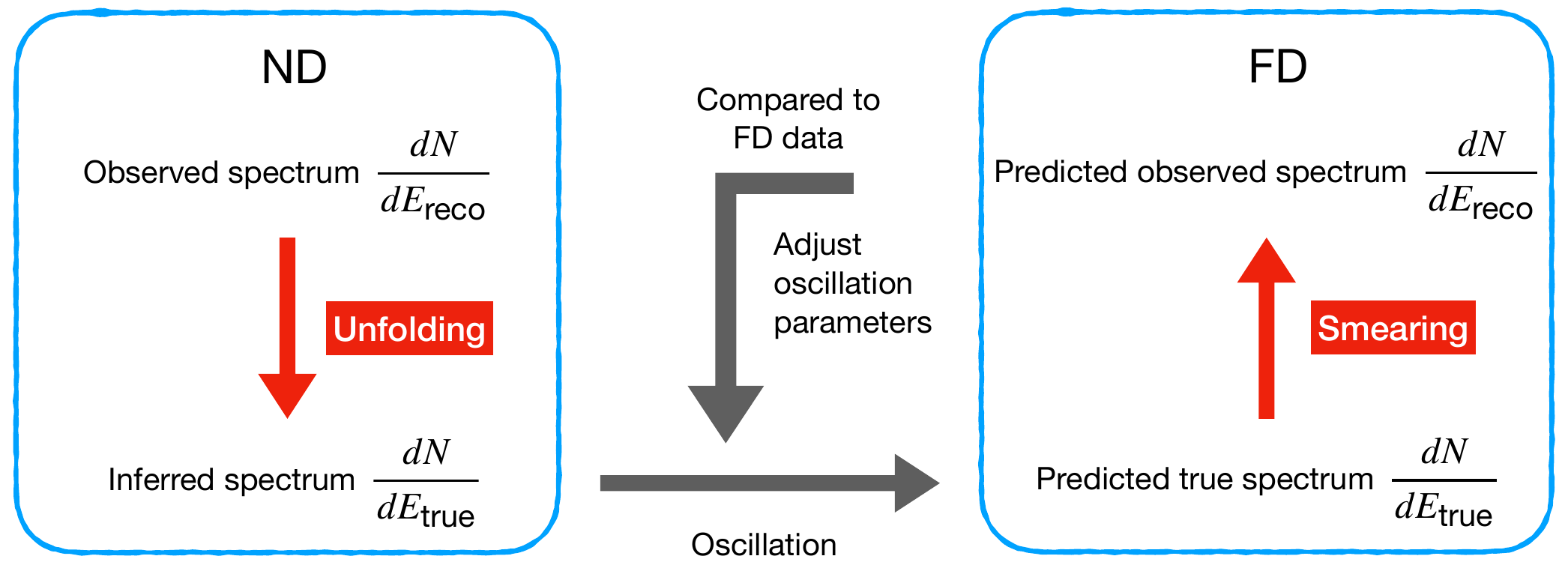}
    \caption{Illustration of NOvA near-to-far-detector extrapolation procedure. The steps utilizing cross section predictions are in red.}
    \label{fig:FD_simulation_prcedure}
\end{figure*}

It is clear from examining the resulting spectra in Fig.~\ref{fig:NDtune} that an initial simulation with a significant deficit in visible hadronic energy has been tuned to a simulation that matches the ND data well in $E_{\mathrm{had}}^{\mathrm{vis}}$. 

\subsection{Near-to-far-detector extrapolation} 
\label{sec:FD prediction}

To further reduce cross section dependence in the oscillation analyses, NOvA employs a data-driven approach to predict FD spectra \cite{NOvA:2021nfi, osti_1827395}. In particular, the FD flux prediction incorporates the unfolded ND flux, rather than relying only on beam simulation.
Figure~\ref{fig:FD_simulation_prcedure} illustrates NOvA's near-to-far-detector extrapolation procedure. It involves the following steps:
\begin{itemize}
    \item Unfold the observed ND data in reconstructed neutrino energy (\enureco) to an inferred spectrum in true neutrino energy (\enutrue). This unfolding employs an ND migration matrix simulated using the tuned generator.
    \item Propagate the resulting spectrum in \enutrue to the FD. This involves the oscillation probabilities as well as factors that account for detector acceptances, solid angles, and other experimental details.
    \item Migrate the predicted FD \enutrue spectrum to a predicted \enureco spectrum using an FD migration matrix simulated using the tuned generator. The predicted \enureco spectrum may then be compared with data.
\end{itemize}

This approach utilizes cross section calculations in the unfolding and migration steps,
which require simulations to model the migration matrices between true and reconstructed neutrino energies. 
Consequently, the predicted distributions of final-state particles and their kinematics become crucial, driving the cross section dependence of the analysis. To illustrate, consider two events with the same true incoming neutrino energy but different interaction types. A deep-inelastic scattering event with multiple knocked-out neutrons and pions may yield a significantly different \enureco compared to a quasi-elastic scattering event with a single outgoing proton. Therefore, the relative modeling of different interaction channels and final-state interactions plays a pivotal role in shaping the analysis outcomes. 

The unfolding step in this analysis can be conceptualized as an inversion problem. 
While various methods exist for unfolding, we employ iterative Bayesian unfolding~\cite{DAGOSTINI1995487} through the RooUnfold package~\cite{Adye:2011gm}. 
This procedure involves an iterative fit process: A test spectrum in true neutrino energy is migrated to reconstructed energy using a generator-based migration matrix.
The resulting spectrum is compared to data using a $\chi^2$ statistic, and the test spectrum is iteratively updated to minimize the difference between the resulting reconstructed spectrum and the observed data.
The process is terminated after a specified number of iterations.
We have tested this unfolding procedure and found that our results are robust against the choice of the unfolding methods.

To propagate the unfolded ND spectrum to the FD, we employ an FD-to-ND ratio that accounts for the different expected fluxes and normalizations in the FD and ND for the respective experiments. For NOvA, the ratio is taken from Fig.~4.16 in Ref.~\cite{osti_1827395}. The oscillations are calculated including matter effects, with a density of $\rho = 2.828$ g/cm$^3$. 

\subsection{NOvA oscillation results}

As a check of the above mock analysis procedure, we examine the results of employing this procedure for the NOvA experiment. Since the NOvA experiment is primarily statistics-limited, we expect that the cross section dependence will not strongly influence the results of the analysis; this is in contrast to DUNE, which will be systematics-dominated. The statistics we use for this mock NOvA analysis match the event rate for $13.6\times10^{20}$ POT in the FHC mode and $12.5\times10^{20}$ POT in the RHC mode~\cite{NOvA:2021nfi}, leading to $\mathcal{O}(100)$ disappearance signal events and $\mathcal{O}(10)$ appearance signal events in the FD for each mode. We scale the untuned simulation generator by an overall factor of 0.9 to approximately mimic the scenario in the NOvA ND tune without overly specifying to their particular spectra; this choice of normalization does not meaningfully impact the resulting best-fit regions. 

We perform an example analysis for simultaneous fits of $\Delta m_{32}^2$, $\sin^2 \theta_{23}$ and of $\delta_{\rm CP}$, $\sin^2 \theta_{23}$.
We assume normal hierarchy and fix the other oscillation parameters to those used in Ref.~\cite{NOvA:2021nfi}: $\Delta m_{21}^2=7.53\times10^{-5}$ eV$^2$, $\sin^2\theta_{12}=0.307$, and $\sin^2\theta_{13}=0.021$. 
Implementing a marginalization over these parameters slightly enlarges the size of the best-fit regions but does not meaningfully affect the conclusions of our analysis.  For flux uncertainties, we employ values of 20\% overall normalization, 4\% relative ND-to-FD normalization, 2\% bin-to-bin uncorrelated, and 2\% correlated bin-to-bin spectral uncertainty.

\begin{figure*}
    \centering
    \includegraphics[width=0.9\textwidth]{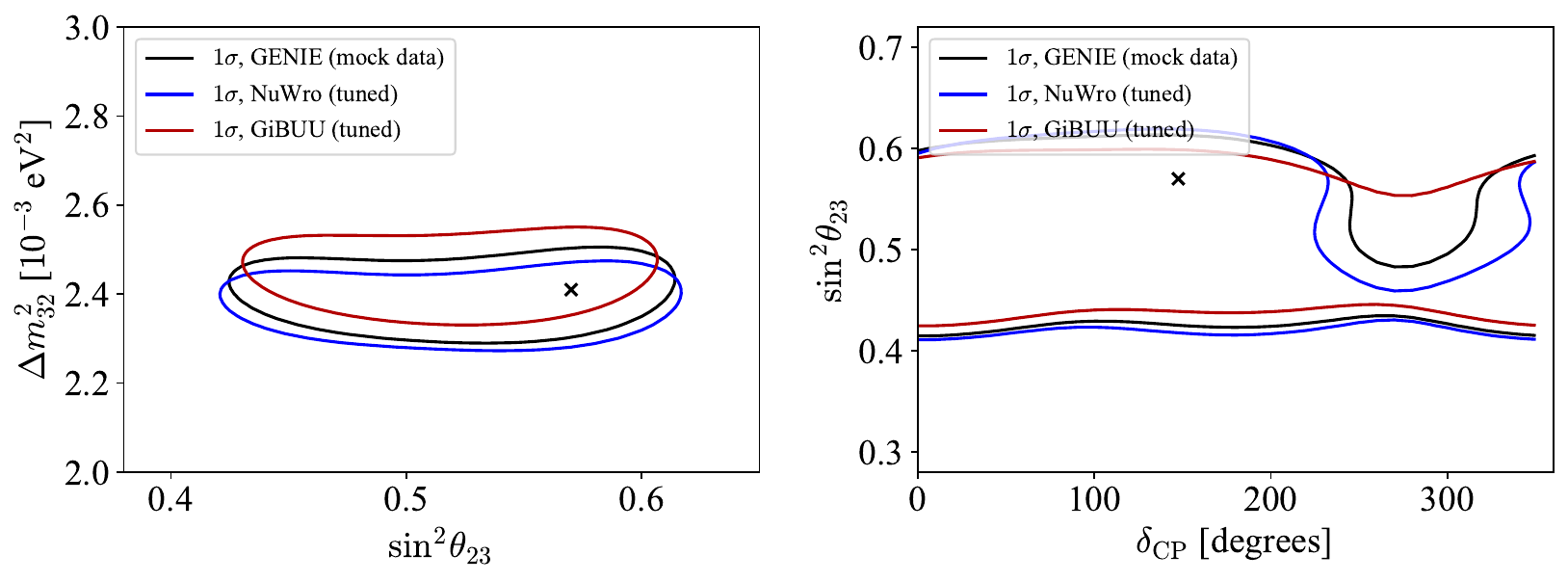}
    \caption{Oscillation analyses for NOvA setup. For all regions, the mock data is generated using GENIE, whereas the analysis is performed with either GENIE, NuWro, or GiBUU. The input truth oscillation parameters are indicated by the black crosses.}
    \label{fig:NOVA_oscillation}
\end{figure*}

Figure~\ref{fig:NOVA_oscillation} shows the results. 
The black region corresponds to performing the full analysis procedure using GENIE; because the mock data is also generated using GENIE, this region represents the case in which one employs a cross section model that accurately models nature. 
The results using NuWro are plotted in blue, while the ones for GiBUU are shown in red. 
Because the regions obtained using the tuned and untuned generators differ only slightly, we only show the tuned ones.

Our analysis yields region sizes and shapes comparable to those reported in the NOvA analyses~\cite{NOvA:2021nfi}, validating our procedure. 
The regions using different simulators vary due to their distinct underlying physics models. 
Notably, for both analyses, the NuWro region closely resembles the ``Truth'' GENIE region, while the GiBUU result shows more divergence. 
This difference likely stems from GiBUU's significantly different nuclear physics modeling, which we discuss in detail in Sec~\ref{sec:DUNE}. 
Despite these differences, even the GiBUU region has substantial overlap with the true region in both panels of Fig.~\ref{fig:NOVA_oscillation}. 
We conclude that the choice of simulation generator does not significantly impact the results of either analysis. 
This finding aligns with expectations, given that the NOvA analysis is predominantly statistics-driven. 


\section{DUNE oscillation analysis} 
\label{sec:DUNE}

We now move to employing this oscillation analysis procedure for DUNE to examine the cross section impact. 

\subsection{Tuning and FD prediction setup}

We make a few changes to the details of the analysis in order to model the DUNE scenario rather than NOvA.
The event files for DUNE are generated for an argon target rather than carbon, and are weighted by the expected DUNE ND flux available at Ref.~\cite{DUNE_fluxes}. 
The ND-to-FD ratio, which accounts for detector sizes and solid angles, is derived from a ratio of the predicted FD and ND fluxes, and is similar to that shown in Fig.~3 of Ref.~\cite{DUNE:2020jqi}. 
Meanwhile, the event rate is scaled to have statistics equivalent to an exposure of 624 kt-MW-years, or 10 years of running with the staging assumptions outlined in Ref.~\cite{DUNE:2020jqi}. 
Following Ref.~\cite{DUNE:2021cuw}, we employ a simple overall normalization uncertainty of 8\% on the $\nu_{\mu} (\bar{\nu}_{\mu})$ events and 5\% on $\nu_e (\bar{\nu}_e)$ events; we have taken slightly larger values than those in Ref.~\cite{DUNE:2021cuw} to approximately account for background uncertainties, which we do not explicitly include. 
We find similar region sizes and shapes to those generated by the GLOBES code in Ref.~\cite{DUNE:2021cuw}.

The ND tuning procedure requires a slightly more meaningful modification. 
While the MEC event distribution covers most of the parameter space in the two-dimensional ND tune plane occupied by the NOvA events, the same is not true for the expected DUNE flux; in particular, a greater contribution from deep-inelastic scattering events leads to a large portion of the parameter space that is not covered by MEC. 
As such, a choice must be made as to which interaction process or set of events to tune. 
For our analysis, we choose to tune the full cross section rather than a specific interaction process.

\subsection{DUNE oscillation results}

\begin{figure*}[ht]
    \centering
    \includegraphics[width=0.9\textwidth]{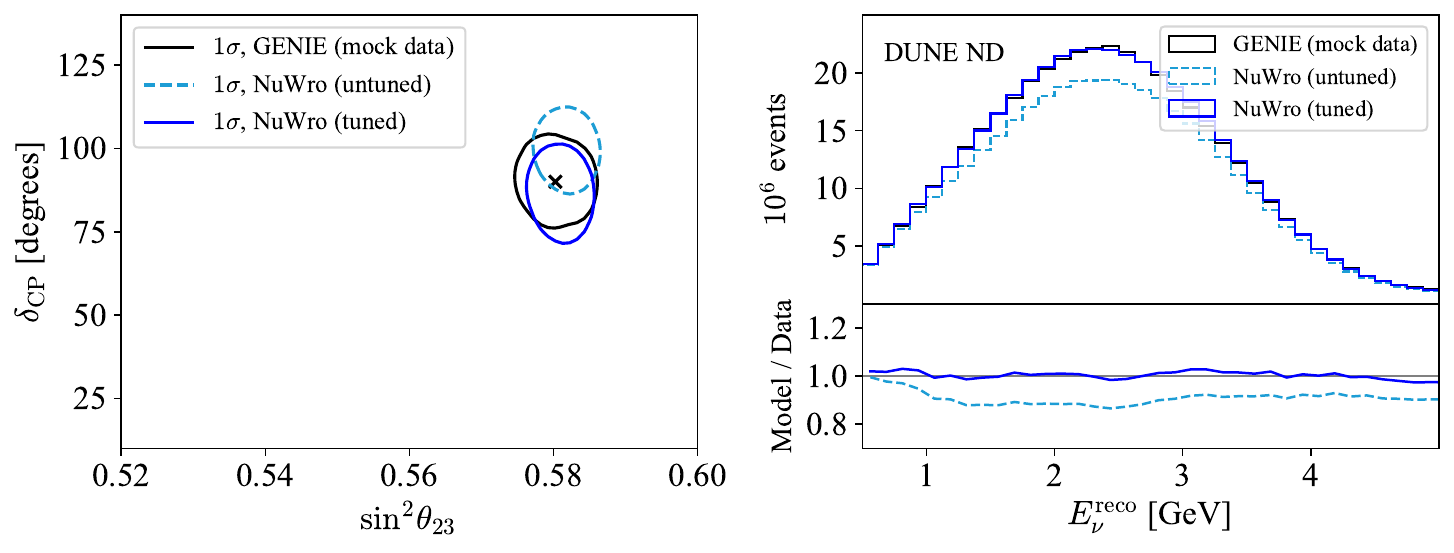}
    \caption{Oscillation analyses for DUNE setup. For all cases, the mock data is generated using GENIE, whereas the analysis is performed with GENIE, both untuned and tuned NuWro. The input parameters are indicated by the black cross.}
    \label{fig:dCP_nuwro}
\end{figure*}

\begin{figure*}[ht]
    \centering
    \includegraphics[width=0.9\textwidth]{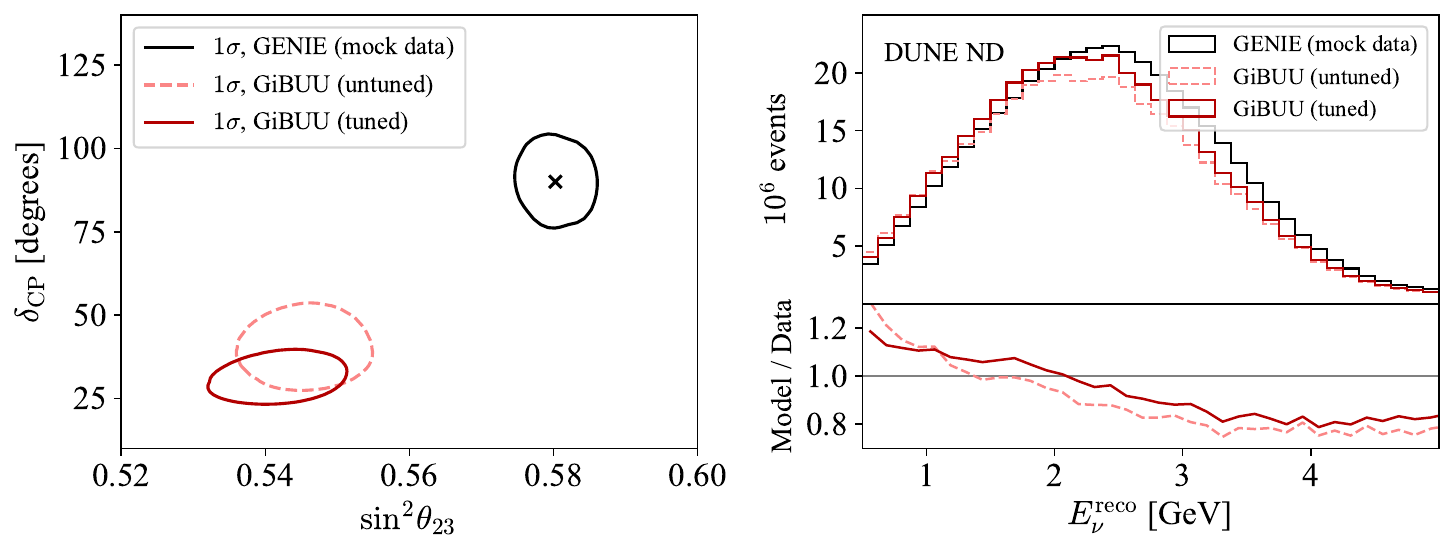}
    \caption{Oscillation analyses for DUNE setup. For all cases, the mock data is generated using GENIE, whereas the analysis is performed with GENIE, both untuned and tuned GiBUU. The input parameters are indicated by the black cross.}
    \label{fig:dCP_gibuu}
\end{figure*}

As before, we generate mock ND and FD data using GENIE, while employing NuWro and GiBUU as simulation generators. 
For each generator, we perform the full analysis procedure, including ND tuning and near-to-far-detector extrapolation using both its tuned and untuned versions. 
The results for NuWro and GiBUU are presented in Fig.~\ref{fig:dCP_nuwro} and Fig.~\ref{fig:dCP_gibuu}, respectively. 
The left panel of each figure shows the 1$\sigma$ region for each of the three generators used in the analysis. 
The right panel shows a comparison of the ND mock data with ND spectra simulated using both the untuned and tuned generators: the upper subplot shows histograms of the spectra, while the lower subplot presents the ratio of simulation to data for both generator versions.

From Fig.~\ref{fig:dCP_nuwro}, we see that employing the NuWro generator to fit GENIE mock data results in best-fit regions that are similar to the truth region. 
While the untuned region shows a slight bias, the tuned region has a significant overlap with the truth region. 
The ND spectra, meanwhile, are in good agreement: although the untuned NuWro generator underpredicts the ND spectrum, the tuned NuWro generator matches the ND data well, and the ratio between the simulation and data varies only slightly around~1.

\begin{figure}
    \includegraphics[width=0.45\textwidth]{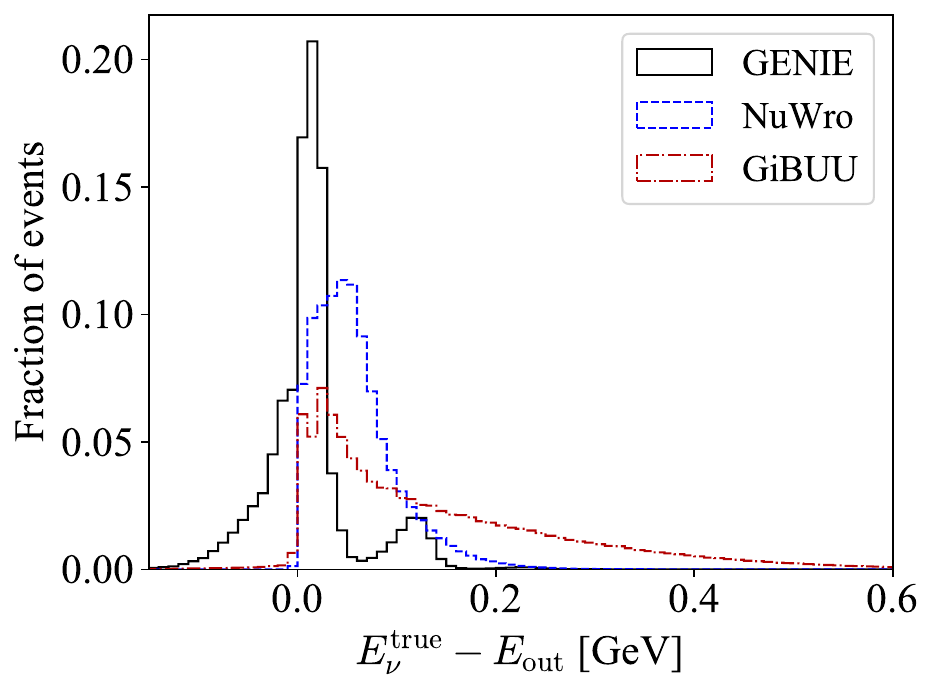}
    \caption{Distribution of difference between true neutrino energy and total available energy in an event, $E_\nu^{\text{true}}- E_\text{out}$.
    The total available energy $E_\text{out}$ includes the total energy of leptons and pions (and other mesons) and the kinetic energy of nucleons.}
    \label{fig:Emiss}
\end{figure}

From these results, one may infer that this oscillation analysis procedure does not strongly depend on the cross section model, and that the tuning procedure can successfully limit what impacts remain. 
However, Fig.~\ref{fig:dCP_gibuu} indicates that this is not the case for all possible implementations of current cross section modeling. 
The untuned GiBUU region shows significant bias relative to the true region; the tuning, meanwhile, does not mitigate this bias,  despite a successful fit to the tune parameters (see Appendix~\ref{sec:appendix} for comparison spectra). 
This illustrates that a naive implementation of the oscillation analysis procedure could lead to biased results for the oscillation parameters.
It is then important to ask if there is a way to detect this potential bias by leveraging specific observables. 
In this case, the right panel makes it clear that the simulation---even after tuning---does not successfully model the observed ND data. 
One may therefore examine the ND spectra to conclude that the tuning procedure fails and that an oscillation analysis employing the tuned cross section model could present strong biases.

\begin{figure*}
    \centering
    \includegraphics[width=0.9\textwidth]{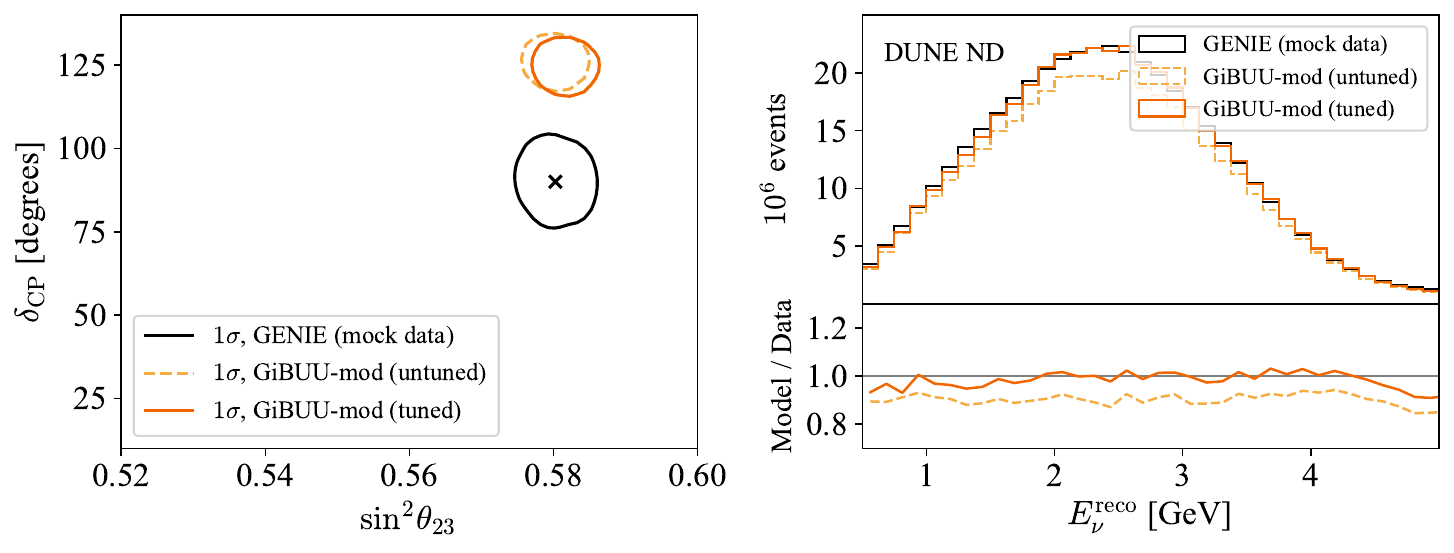}
    \caption{Oscillation analyses for DUNE setup. For all cases, the mock data is generated using GENIE, whereas the analysis is performed with GENIE, both untuned and tuned GiBUU-mod (our phenomenologically modified GiBUU). The input parameters are indicated by the black cross.}
    \label{fig:deltaCP_gibuuprime}
\end{figure*}

There are many differences in how individual event generators implement the primary cross sections and the nuclear effects. 
However, given that cross section models affect the analysis through migration matrices, a potential reason that GiBUU results are so different from GENIE and NuWro is how final-state interactions (FSI) are implemented. 
With the impulse approximation, after the primary neutrino-nucleon interaction, the secondary hadrons still need to propagate inside of the nucleus from their production point to where they exit the nucleus. 
During propagation, the secondary hadrons could scatter with other nucleons in the nucleus, producing new particles or transferring energy. 
GENIE 3.04.00 (with G18\_02a\_00\_000 tune)  and NuWro 21.09.2 treatments of FSI are similar in that both employ hadronic transport models. 
In these models, nucleons propagate as ``billiard balls'' throughout the nucleus (or some simplified version of this), possibly scattering off other nucleons according to nucleon-nucleon cross sections (see e.g. Refs.~\cite{Boudard:2002yn, Andreopoulos:2009rq, Golan:2012wx}).
In-medium effects and the nucleon configuration may differ between these generators, but the overall physics is similar. 
GiBUU, on the other hand, solves the quantum kinetic Boltzmann equations, also known as Kadanoff-Baym equations~\cite{KadanoffBaym:1962,Botermans:1990qi}.
Since obtaining exact solutions is currently not possible, GiBUU
adopts the gradient approximation, which assumes the nuclear density distribution to be fairly flat.

To be quantitative, we show in Fig.~\ref{fig:Emiss} the distribution of the difference between true neutrino energy and total available energy in outgoing particles, that is $\Delta \equiv E_\nu^{\text{true}}- E_\text{out}$.
The total available energy $E_\text{out}$ includes the total energy of leptons and pions (and other mesons) and the kinetic energy of nucleons. 
As we see in Fig.~\ref{fig:Emiss} the distribution of $\Delta$ is similar for GENIE and NuWro, while GiBUU predicts a wider distribution due to the different modeling of FSI. 
Therefore, when inferring the neutrino energy from a given calorimetric energy, GiBUU would predict higher neutrino energies compared to GENIE and NuWro.
Note that this difference $\Delta$ for monochromatic neutrinos from kaon decay-at-rest has been studied experimentally by the JSNS$^2$ experiment, and it was found that GiBUU predicts a broader $\Delta$ distribution than data, while NuWro presents a better fit~\cite{Marzec:2024fka}. Motivated by Fig.~\ref{fig:Emiss}, Gallmeister and Mosel have further investigated this wide distribution in Ref.~\cite{Gallmeister:2025iug}.

To test our hypothesis and further evaluate the range of biases generators can give, we examine a phenomenologically modified version of GiBUU, denoted by GiBUU-mod, which enforces $\Delta$ closer to zero values. 
We add the difference $E_\nu^{\text{true}}- E_\text{out}$ to the hadronic energy, with a 5\% smearing included on this contribution when calculating the visible hadronic energy. 
This gives us an example generator more similar to GENIE and NuWro in terms of this energy difference, while still using a completely different FSI framework.

Figure~\ref{fig:deltaCP_gibuuprime} shows the results of this analysis. 
As we can see in the right panel, this modified version of GiBUU provides a modeling of the ND mock data after tuning that is closer to GENIE.
However, the relationship between true and reconstructed energy still does not match the one used in generating mock data, as there remains a bias in the preferred regions in $\delta_{\rm CP}$. 
This seems to indicate that the ND data is not sufficient to constrain the cross section models accurately, and the oscillation analysis could be biased outside of the error budget.

\begin{figure}
    \centering
    \includegraphics[width=0.45\textwidth]{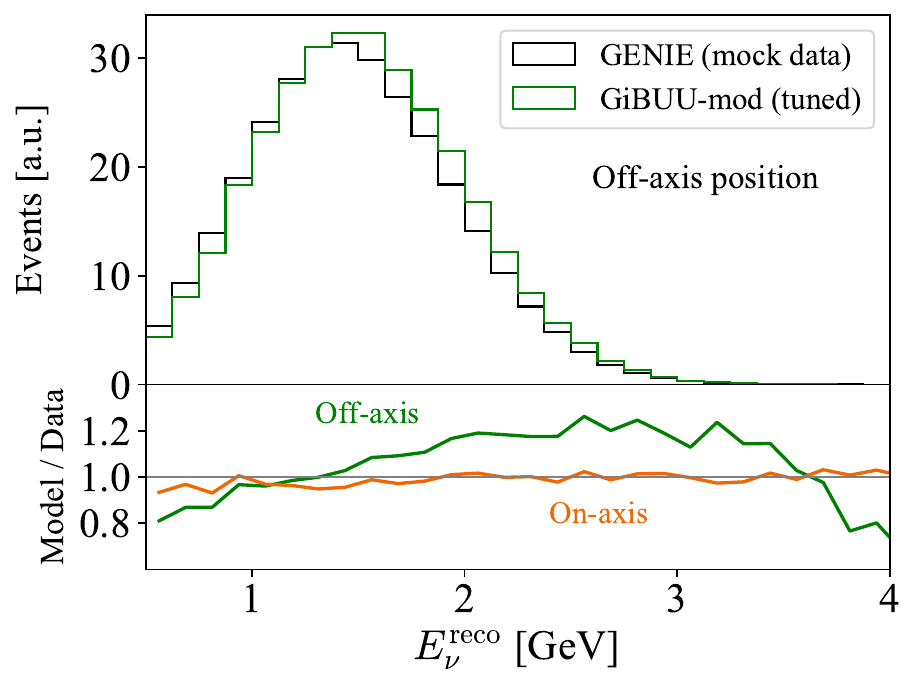}
    \caption{An illustration of the power of DUNE-PRISM. In the top panel, we show the ND spectra for an example off-axis flux for the data and the tuned generator. In the bottom panel, we show the ratio of the spectra in green. The orange line shows the ratio for on-axis spectra, taken from Fig.~\ref{fig:deltaCP_gibuuprime}.}
    \label{fig:gibuuprime_offaxis}
\end{figure}

Although this may be worrisome at first, we note that more experimental handles can be used to flesh out the discrepancies even in this case.
DUNE near detector will be movable to probe different off-axis angles, a technique coined DUNE-PRISM~\cite{DUNE:2021tad}.
By probing different angles with respect to the neutrino beam axis, DUNE-PRISM can probe different neutrino spectra correlated by the pion energy spectrum.

We show an example off-axis spectrum in Fig.~\ref{fig:gibuuprime_offaxis}, where the flux has been pulled from a simple Gaussian distribution centered at $E_{\nu}^{\mathrm{true}}=1.5$ GeV with a width of 0.5 GeV. 
We do not employ a specific prediction for the number of events relative to the nominal on-axis flux, instead focusing on the relative shapes of the predicted spectra. 
Although the on-axis post-tune \enureco spectrum agrees well with the mock GENIE data, the off-axis flux presents a large discrepancy.
This is due to the different relationships between \enutrue and \enureco in GENIE and GiBUU-mod. 
Our findings are similar to those found by DUNE in the TDR~\cite{DUNE:2020ypp}.

This result suggests that more off-axis ND data, using the DUNE-PRISM technique, is crucial for breaking the degeneracies in specific experimental observables and providing potentially faithful constrained cross section models. 
Investigating how well this works quantitatively requires detailed DUNE-PRISM information and analysis strategies that are not publicly available. We, therefore, leave it to future work for quantitative studies.


\section{Conclusion} 
\label{sec:conclusions}

In this work, we studied in detail the impact of neutrino-nucleus cross section modeling on the robustness of long-baseline oscillation experiments. 
We have considered three neutrino-nucleus event generators to capture the current cross section uncertainties. 
We then performed mock oscillation analyses with strategies inspired by NOvA's procedure, which leverages near detector data to mitigate uncertainties, and considered setups similar to both NOvA and DUNE.

We find that the NOvA analyses are largely unaffected by biases arising from cross section modeling due to limited statistics, which is consistent with official results~\cite{NOvA:2021nfi}. 
However, DUNE sensitivity could be significantly biased due to mis-modeling. 
Extreme biases arise when we simulate data using GENIE and perform the analysis using GiBUU; this is due to the drastic differences in the neutrino-nucleus interaction models.
Nevertheless, one can identify these issues by examining \enureco spectra after tuning and observing whether discrepancies remain. 
However, in other cases such as when we perform the analysis using a phenomenologically modified version of GiBUU, the ND spectra match well for on-axis fluxes, although biases in the oscillation analysis remain.
To combat this scenario, utilizing off-axis ND data with DUNE-PRISM is crucial. 
We leave quantitative studies along this line for future work. 

Our results illustrate a need for improvements in the theoretical cross section modeling, as well as new experimental approaches to address cross section uncertainties.
One way of testing the tuning procedure would be to perform experimental analyses with several event generators.
Measurements of $\nu$-argon cross sections from experiments such as SBND may contribute to the input cross section models employed in future DUNE analyses. 
At the same time, DUNE-PRISM off-axis ND data can provide further constraints on cross section modeling. 
A procedure that employs these and other strategies will be necessary to ensure the success of future long-baseline neutrino experiments.


\begin{acknowledgements}
We thank Jianming Bian, Joshua Isaacson, Kevin Kelly, Joachim Kopp, Ulrich Mosel, Afroditi Papadopoulou, Xin Qian, Linyan Wan, and Daniel Whiteson for helpful discussions. 
The work of N.C. is supported by the US National Science Foundation under Grant PHY-2210283.
This manuscript has been authored by Fermi Forward Discovery Group, LLC under Contract No. 89243024CSC000002 with the U.S. Department of Energy, Office of Science, Office of High Energy Physics.

The work and conclusions presented in this publication are not to be considered as results from the DUNE collaboration.
\end{acknowledgements}


\newpage
\appendix


\section{Tuned spectra} 
\label{sec:appendix}

In this Appendix, we present the results of the tuning procedure for two illustrative cases: tuning NuWro to data generated with GENIE, and tuning GiBUU to data generated by GENIE. We use event files simulated for neutrino interactions on argon. In both cases, we find that while there exists significant disagreement between the mock data and the untuned simulation, the tuned generator fits the data well in both observables.

\begin{figure*}
    \includegraphics[width=0.45\textwidth]{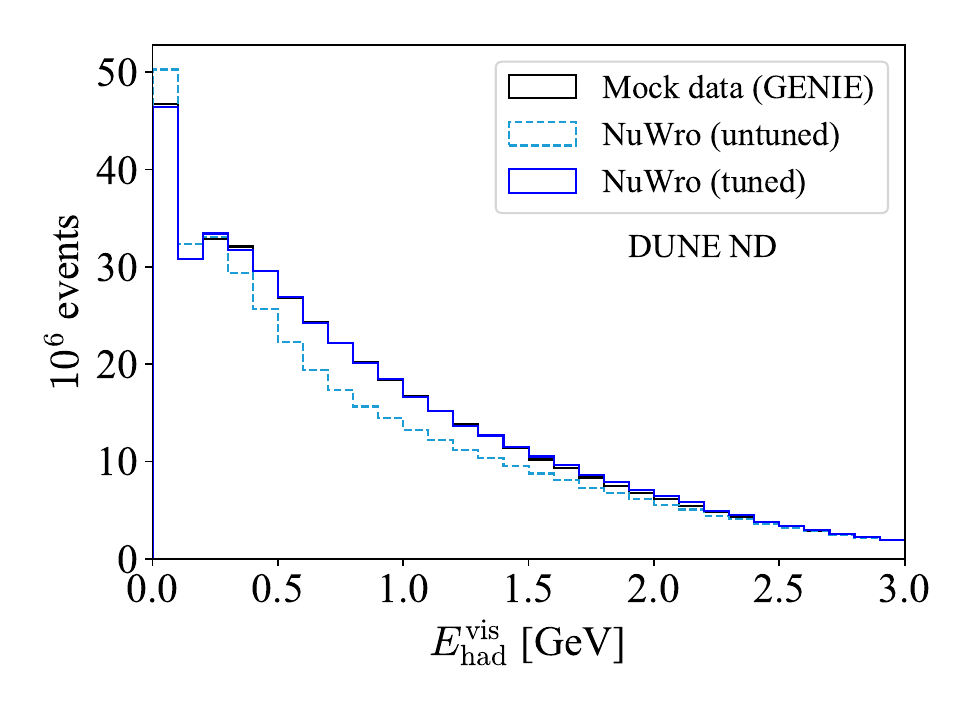}
    \includegraphics[width=0.45\textwidth]{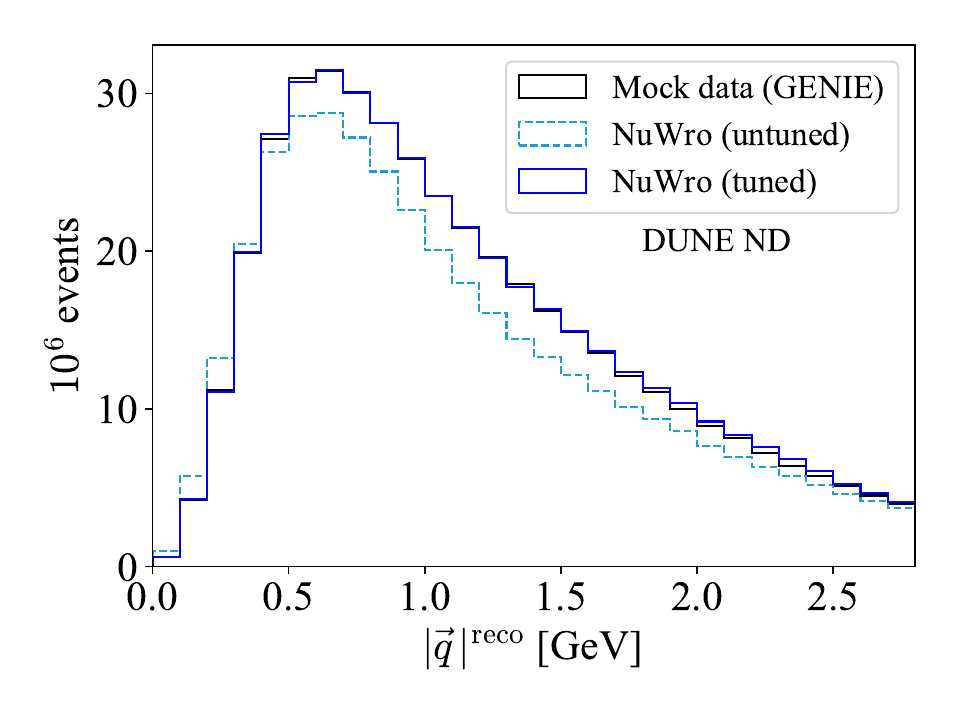}
    \caption{Comparison of spectra in $E_{\mathrm{had}}^{\mathrm{vis}}$ (left) and \qreco (right) for GENIE mock data, untuned, and tuned NuWro.}
    \label{fig:nuwro_tune}
\end{figure*}

\begin{figure*}
    \includegraphics[width=0.45\textwidth]{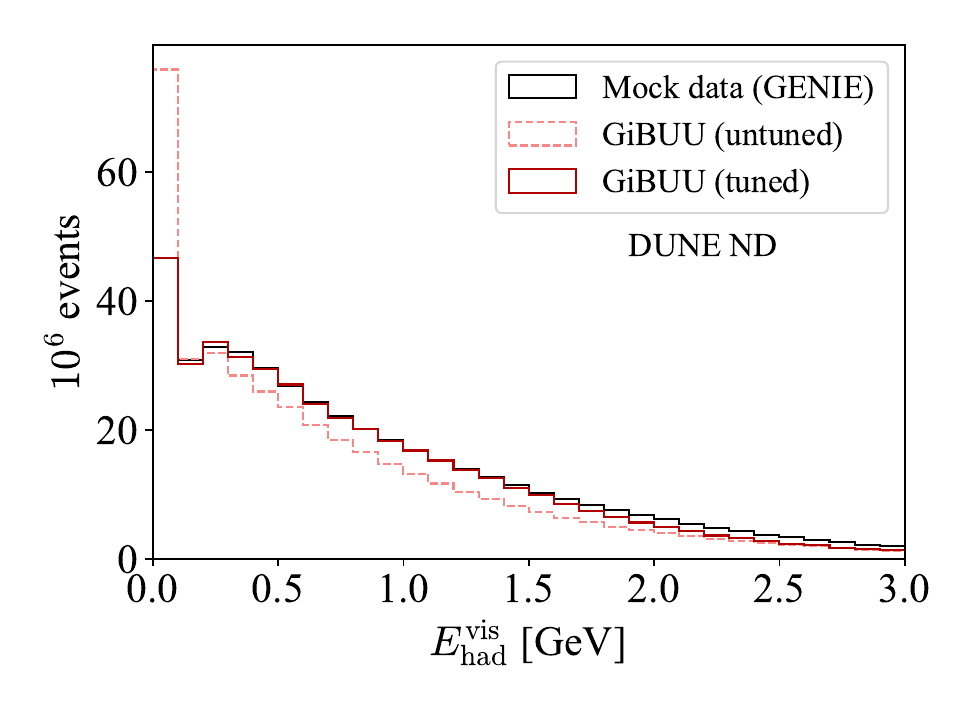}
    \includegraphics[width=0.45\textwidth]{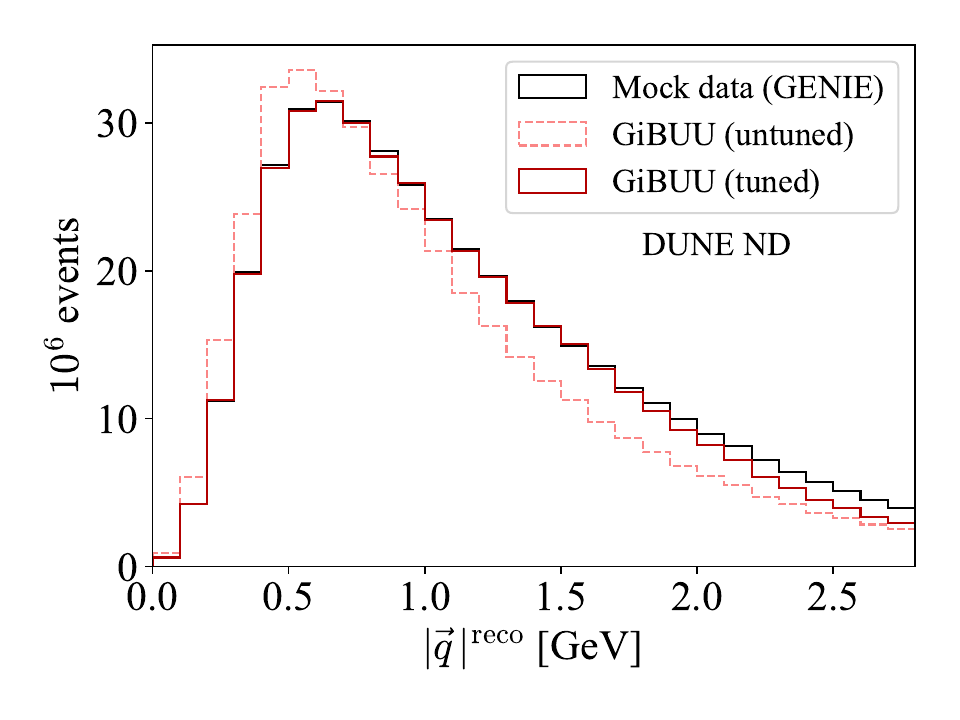}
    \caption{Comparison of spectra in $E_{\mathrm{had}}^{\mathrm{vis}}$ (left) and \qreco (right) for GENIE mock data, untuned, and tuned GiBUU.}
    \label{fig:gibuu_tune}
\end{figure*}

%


\appendix

\end{document}